\documentstyle[preprint,aps,epsf,prbbib]{revtex}
\tightenlines
\begin{document}
\draft
\title{Energetics and bonding properties of the Ni/$\beta$-SiC (001) interface:
 an {\it ab-initio} study}

\author{G. Profeta, A. Continenza}
\address{
Istituto Nazionale di Fisica della Materia (INFM)\\
Dip. Fisica,
Univ. degli Studi dell'Aquila, 67010 Coppito (L'Aquila), Italy }
\author{A. J. Freeman}
\address{Department of Physics and Astronomy \\
Northwestern University, Evanston, IL 60208 (U.S.A.)\\}
\maketitle

\begin{abstract}
We investigate the adsorption of a Ni monolayer on the $\beta$-SiC(001) surface  
 by means of highly precise first-principles all-electron FLAPW calculations. 
 Total
energy calculations for the Si- and C-terminated surfaces reveal high
Ni-SiC adsorption energies, with respect to other metals, 
confirmining the strong reactivity and the stability
of the transition metal/SiC interface.  These high binding
energies, about 7.3-7.4 eV, are shown to be 
related  to  strong $p$-$d$ 
hybridization, common to both surface terminations and  
different adsorption sites and despite the large mismatch, can stabilize
overlayer growth. A detailed analysis of the bonding mechanism,
hybridization of the surface states,
charge transfer and surface core level shifts reveals the strong 
covalent character of the bonding. After a proper accounting  
 of the Madelung term, the core level shift is shown to follow the
 charge transfer trend.    
\end{abstract}


\section{Introduction}
The promising mechanical and electronic properties of silicon carbide are
stimulating many studies focused on the application of its semiconducting 
and extreme structural properties. In fact, the interest towards SiC is twofold:
 on one hand SiC is a high-strength composite and   
 high-temperature ceramic seal\cite{bohn} and is considered  a basic 
 material for detectors and
devices operating in high radiation environments, in space applications, 
in medicine and industry. In this regard, as pointed out in recent
 papers \cite{Hoek,Hoek2}, fundamental interest is focused on 
 the investigation of the nature of bonding and the 
 origin of adhesion at metal-ceramic interfaces: the bonding properties, 
 are in fact strongly dependent on the metal and on the ceramic and may lead 
 to nonreactive systems\cite{freem} (e.g. MgO/Ag) as well as
 to reactive systems (e.g. SiC/Al).
On the other hand, SiC represents an interesting semiconductor.
It has a wide band gap (2.2 eV); carrier mobilities and drift velocities 
exceed the corresponding 
values in Si by more than a factor of two. Together with 
its thermal stability, these characteristics allow for high-power and 
high-frequency operation at 
elevated temperatures \cite{hoc}. However, application of 
SiC-based
semiconducting devices requires a deep understanding  of metal contact 
formation. 

As in the case of other semiconducting
materials,  knowledge of the reactivity and of the detailed electronic structure 
of the
metal-semiconductor interface is the first step towards device design  and
performance control.
An excellent review by Kaplan and Bermudez \cite{berm} presents results 
of metal deposition on various SiC surfaces obtained by different experimental 
techniques.
Among the metals studied, Ni seems to have the highest stable temperature 
contacts,
an essential requirement to exploit the properties of SiC: Ni overlayers were 
reported to be thermally stable up to $ 800 ^{\circ}$ C; 
at higher
temperatures the silicide formation process starts leading to a silicide-SiC 
contact. 
During heat treatment, the formation of silicides with different
stoichiometry was also reported.  
The silicidation process, that is metal deposition followed by rapid thermal 
annealing (RTA), is very common in silicon-based technology and
many experimental and theoretical works were  devoted to
understanding this process and, in particular, to characterize the multiple 
silicide phases and their interface with Si. We cite, specifically, 
the recent and extended work by the Vienna group \cite{raimound}, which 
studied the cohesive, 
electronic and structural properties of various TM-silicides (TM=Fe, Ni, Co), 
including different bulk phases, surfaces and interfaces with Si. Looking at 
the epitaxially
stabilized structures of Fe, Co, and Ni silicides they also discovered new 
artificial materials with interesting elastic properties that they called 
{\it supersoft}. As an example, FeSi$_{2}$ in the C1 (fluorite) structure, as
well as  CoSi and
NiSi in the B2 (sodium cloride) structure, show zero strain energy under biaxial
strain on a (100) substrate. However, these phases are  not stable phases and 
the lattice mismatch with the Si(100) surface is too large to
encourage epitaxial growth.

As is well-known,
pseudomorphic epitaxial growth of  ultrathin metal films on a substrate results
in a strain on the film  due to the mismatch
between the lattice constant of the film and the substrate. Especially 
interesting is the stabilization of metastable phases that can be achieved using
suitable substrates, offering strong adhesion energy and close 
matching conditions. 
Within this framework, the investigation of the Ni-SiC(001) interface becomes 
particularly relevant 
in view of the study of a possible Ni-silicide/SiC(100) system. Moreover, 
from an
experimental point of view, the study of this interface is far from the 
deep understanding achieved for the silicide-silicon interfaces, and in fact
the real structure of the interface is still unknown. 
Finally Ni/SiC interface, deserves attention in its own right, since this rather simple system
can provide important information on the nature of the transition-metal SiC
bond.
For all these reasons, we present here the first {\it ab-initio}
 analysis of the Ni/SiC 
interface at the first stage of deposition, focusing on the system
formed by a Ni monolayer  adsorbed on a SiC(001) substrate.

The paper is organized as follows: in Sect. \ref{tecni} we give details of
the calculation, then we discuss results for the structural  (Sect. \ref{str})
and electronic properties (Sect. \ref{ele}) for all the different geometries
considered; and finally, in Sect. \ref{cori}, we discuss the surface core-level
shifts and draw conclusions (Sect.\ref{concl}).
\section{Technical details}
\label{tecni}
The calculations are performed  using the all--electron full-potential 
linearized augmented plane wave (FLAPW) method \cite{flapw} 
  within density functional theory in the local density approximation (LDA) 
  \cite{dft} according to the
 Hedin-Lundqvist parametrization of the exchange-correlation energy. 
 The surface is modelled in a superslab
 geometry with a slab that includes seven (six) silicon layers and six (seven)
  carbon layers to simulate the substrate in the case of the silicon (carbon)
  terminated surface with a Ni adsorbate on both surfaces and 10.5 \AA \
  (equivalent to about ten SiC bilayers) 
  for the vacuum region. The calculated bulk SiC equilibrium lattice constant 
  $a_{SiC}=4.35$ \AA\  agrees (within 0.2 $\%$) with the 
  experimental value.
   The muffin-tin radii
  used are $r_{mt}^{Si}=1.8$ a.u., $r_{mt}^{C}=1.5$ a.u. and 
$r_{mt}^{Ni}=2.0$ a.u.. The Ni 3$d$ and 4$s$, Si 3$s$ and 3$p$, C 2$s$ and 2$p$
were treated as valence electrons. We used a plane wave cutoff 
$k_{max}=3.8$ a.u. for the  basis set, 8 a.u. for the potential and charge 
density representation and a 6 $\times$ 6 $\times$ 4 Monkhorst-Pack \cite{MP}
$k-$point mesh for integration over the Brillouin zone; 64 $k$ points in the
irreducible wedge of the Brillouin zone were 
used to  
calculate the density of states.  Our present implementation
includes total energy and atomic force calculations, 
 which allows full structure optimization  using  a {\em quasi}--Newton 
 \cite{broy} scheme.
 The stable configuration is found
when the 3{\em n}-dimensional force vector of the system with {\em n} atoms
 is approximately zero (i.e. forces on each atom  smaller than 1.0 mHtr/a.u.).
For all the structures considered, we used the same supercell to
achieve consistency for the total energy. 
Convergence, within 1 mHtr in the total energy was checked as a function of 
k-points integration, $k_{max}$ value
and number of layers used in the calculation.

\section{ Structural properties of a Ni monolayer on SiC(100)}
\label{str}
A top view of the unit cell adopted for the simulation is illustrated in 
Fig.\ref{struct}.
The most stable reconstruction for the clean Si-terminated (Si-SiC)(001) 
surface is calculated to be p(2$\times$1) with unbuckled Si dimers 
\cite{catell} which has a 
total energy only a few meV/atom lower than the ideal p(1$\times$1).
 For the C-terminated (C-SiC)(001) surface, the most likely stable 
 reconstructions
 are  c(2$\times$2) and p(2$\times$1) \cite{catell2}.
 
 With only few exceptions, materials with low surface energy tend to grow  on
  high-surface-energy materials, while the reverse is usually inhibited.
  Ni has a high surface energy, about 0.96 eV/atom, a value much higher than
  in non-transition
  metals and about 2/3 of the upper limit for transition metals (Re, Os)
  \cite{picke}. In a crude approximation, neglecting the interaction 
  energy
  between substrate and overlayer, Ni growth should not be favored when
   compared to other low surface-energy metals. As anticipated in the
 Introduction, however, transition metals with partially occupied $d$ 
 shells show large
 interaction energy with reactive ceramics that may stabilize overlayer
 growth. We demonstrate here that this is the case for the Ni/SiC interface. 

For this particular interface no information, to our knowledge, is available 
regarding favored  adsorption sites and  possible surface 
reconstructions: deposition of Ni on SiC(001)
leads to  Ni-Si intermixing at sufficiently high
temperature and no experimental studies are available on the non-intermixed 
Ni-SiC
phase. Due to the lack of experimental information, our calculations on the
Ni-SiC(001) interface assume a (1$\times$1) in-plane 
periodicity. This system allows the study of the bonding properties of the
transition metal (TM) on both the Si-terminated and C-terminated
SiC surface and gives insights into possible
geometries and more complicated reconstructions.
  In the ideal (1$\times$1) reconstruction, in fact, there are four possible 
  adsorption sites (Fig \ref{struct}): antibridge, 
bridge, four-fold (hollow) and on-top sites that constitute total energy
extrema, due to the surface symmetries.
From a geometrical point of view, the four adsorption sites are rather different
and hence different bonding properties of the TM atom can be expected. For each 
site, we
calculated the equilibrium structure and the adsorption energy, i.e. the energy
gained in the adsorption process, which we define as:
$$ E_{ads}=-{1\over{2}}(E_{s+ad}-E_{s}-2E_{ad})$$
here, $E_{s+ad}$ is the total energy of the SiC-Ni system, $E_{s}$ is the total
energy of the clean surface  and $E_{ad}$ that of the free Ni atom.
Results for the Si and C-terminated surfaces are reported in 
 Table \ref{param} (we do not report values for the on-top site since it shows
 very low adsorption energy and is therefore not favored).

As far as the substrate is concerned, all the geometries considered lead to
a small  
relaxation of the upper SiC layer (only in the hollow-site case does the 
outward
relaxation of the Si and C layer reach about 0.1 \AA).
The height of the adatom on the substrate increases in going 
from the hollow site (the most open structure) to the antibridge (the most
closed), in both terminations. In
particular, in the hollow site  Ni is very close to the surface plane, 
about 1 \AA\   in the Si-terminated and 0.5 \AA\ for the C-terminated case. 
The four-fold adsorption site   
was first proposed \cite{hollo} for Co/Si(001); in this case the 
adatom height with respect to the substrate was inferred to be $\simeq 0$, 
implying that Co  would lie in the topmost  
silicon plane restoring the (1$\times$1) reconstruction. This picture is 
consistent with the
results we find: the larger in-plane lattice constant of Si(001), with respect 
to SiC, could
favour penetration of the adatom into the surface, leading to an adatom height 
 lower than in the SiC case. 
 
 The difference observed between the Si and C 
terminations is a consequence of the smaller atomic radius of C with 
respect to Si, while the adsorption energies are
similar (within $\simeq$0.3 eV) in all the systems investigated (see
Table \ref{param}). 
This result is interesting since the adsorption sites considered and the
two terminations are quite different.  
In the hollow site, the Ni atom  is
four-fold coordinated with silicon (carbon), while 
in the bridge site the coordination is only two-fold. On the other hand, 
the antibridge site, which is also two-fold coordinated, has to form bonds with
the dangling-bonds from the two 
top-most surface atoms which point towards the bridge site. 
This reduces the
adsorption energy and makes the antibridge unstable compared to the 
bridge site.
We should remark, however, that substrate relaxation plays an important role: in
fact, if we keep the substrate unrelaxed, we find that the configuration with
the adatom in the bridge site at its equilibrium position has higher
energy (about 1 eV) with respect to the completely relaxed structure for the
hollow site.

As a general remark, we point out that the adsorption energies calculated for 
the three sites considered are all larger with respect to those calculated 
in the case of an Al adatom in the same configurations \cite{al}. 
The computed values for the adsorption energy are referred to a clean,
unreconstructed but relaxed SiC surface and a free Ni atom.
In order to  estimate the binding energy of a Ni monolayer 
on the SiC surface, we calculated 
the total energy of an unsupported Ni monolayer at the SiC lattice constant. 
The total energy difference per atom between a
free Ni atom and a  Ni free-standing monolayer in a square lattice is  
1.5 eV/atom, showing that
in the choosen configuration the Ni-Ni bond-energy is smaller than the 
Ni-Si, Ni-C. The high adsorption energies found indicate that 
Ni/SiC(001) belongs to the chemisorption
class ($E_{ads} >0.5 $ eV) instead of the physisorption class ($E_{ads} <0.5 $ eV),
 thus confirming that Ni/SiC is a reactive metal-ceramic interface where 
 bond hybridization plays an important role.
In particular, it is straigthforward to identify the $p$-$d$ hybridization,
characteristic of TM-silicides and carbides, as having major responsibility 
for the bonding. 
The same bonding features are expected to be found 
for Fe and Co (that also form silicides and carbides \cite{berm,haglu}), as 
discussed in greater detail below.
\section{Electronic properties}
\label{ele}
The shape of the energy landscape of Ni on SiC appears quite flat for the 
Si-terminated
surface, and in general the high adsorption energy of the bridge and hollow
sites deserves a discussion of the electronic properties of both 
configurations. 
In the following, we discuss the electronic and bonding properties for the Ni 
adatom in both sites separately. 
\subsection{Bridge Site}
In this geometry the Ni adatoms occupy the cation (anion) sites and therefore
are in line with the Si (C) $sp^{3}$
dangling bonds. In the clean Si-SiC surface, 
the surface states originate from the surface dangling bonds and lie above the 
bulk valence band maximum, while  for the C-SiC
surface in the $(1\times 1)$ reconstruction, the surface states appearing 
in the bulk band gap are contributed by dangling-bonds as well as back-bonds
of the surface C atoms \cite{catell}. We expect that the interaction with Ni adatoms should
partially remove these states.
In Fig. \ref{dos_bridge_si_c_term}  we show the projected density of states 
(PDOS) on 
different atom sites for the Si-terminated and C-terminated surface resolved
according to surface symmetry,
 from the inner "bulk" layer (Si or C) to the Ni adatom. 
 In the figures shown, we set the
 zero of the energy axis to the Fermi level that, consistent with other 
 metal-semiconductor interfaces, lies within the 
 semiconductor band gap even at monolayer coverage. 
 
 The  two central layers (lower panel in Fig.\ref{dos_bridge_si_c_term}) 
 of the SiC slab are quite representative of bulk SiC, showing 
 a well defined valence band maximum and a shape which closely resembles   
 the PDOS of  bulk
 SiC. The interaction with the Ni adatom is localized within the topmost 
 surface layers. 
 Let us first discuss the Si-terminated surface. As already known from studies
 on TM silicides \cite{TM}, bonding
between TM and silicon atoms is due to hybridization of the Si-3$p$ 
with the  TM $d$ states; in the present case, we also find a partial contribution 
from Si-$s$ states, coming from rehybridization of the original
$sp^{3}$ orbitals. In the same figure, we show the PDOS for the 
Ni  $d$ states resolved into atomic-like $d$-components. 
As a general feature, present in all the structures considered, we found a
noticeable broadening of the Ni-$d$ bandwidth (about 5 eV) when compared to the
narrow $d$-band of the Ni monolayer ($\sim $ 1 eV). 

The bonding properties are now quite clear: there is a 
strong hybridization
between the Si $p$ states (mainly with $p_{x}, p_{z}$ character) with some of 
the Ni $d$-states with mainly $d_{xz}, d_{3z^{2}-r^{2}}$ character with a spatial
distribution pointing directly towards the Si atom dangling bonds: as a result,
 these states have a larger band-width and are pushed to higher 
binding energy 
with respect to the other $d$ states that remain in a non-bonding configuration. 
On the surface Si layer  we find
the presence of a pronounced feature with $p_{x}$-character  above 
(by $\simeq +1 eV$)
and below (by $\simeq -3 eV$) $E_{F}$ and a broadening of the $s$
band. 

Evidence of the character of these states 
comes from the plot of the charge density (shown in Fig. \ref{bond_ant}) 
due to eigenfunctions whose eigenvalues belong to the two main 
peaks (the bonding one at about -3 eV and the
antibonding one at 1 eV). While the charge
density related to the bonding peak (Fig. \ref{bond_ant}(a)) is well 
 delocalized among  Ni and Si atoms (with some resonance states 
into the bulk), 
 the anti-bonding peak (Fig.\ref{bond_ant}(b)) shows very low charge 
 density 
 in-between the Si and Ni 
 atoms with  a horizontal shape characteristic of non-bonding $p_x$ states, 
 well localized in the surface layer.

Analogous bonding features can be found in the C-terminated surface: in this
case, however, the bond involves C-$p$ orbitals combinations with  
mainly  Ni $d_{xz}$ and $d_{x^2-y^2}$  states showing a quite large 
bandwidth in the PDOS;
the C-$1s$ are found at higher binding energies (not shown in Fig.
\ref{dos_bridge_si_c_term}). The other $d$-orbitals 
(i.e. $d_{xy}$, $d_{yz}$, $d_{3z^2-r^2}$) are
confined to a more limited energy range around -1 eV below $E_{F}$.
Correspondingly, we find that the $p$-states of the C-surface site are smeared
 over the same energy range of the Ni-$d$ states leading to a total bandwidth 
 of 7-8 eV.
Again, we can distinguish bonding-antibonding features whose energy
position is clearly symmetry dependent. In fact, we find a large splitting  
(5 eV) for the $p_x$ states hybridized with Ni-$d_{xz}$ (cf. strong feature 
at -4 eV) 
that makes the antibonding combination completely empty, while the
bonding-antibonding  splitting is
only of 3-4 eV for $p_y$ and $p_{z}$ states. 

To further ascertain the 
difference with the
Si-terminated surface, we consider the surface geometry in the two cases. 
While Ni sits at 1.52 \AA\ above the Si-terminated surface, it is almost 
coplanar with the C-topmost layer. This draws more spectral weight into the 
$p_{x}$ orbital which better fits the Ni $d_{xz}$ and $d_{x^2-y^2}$
orbitals.
The bonding features observed for the $p_{z}$ and $p_{y}$ states are explained 
recalling that states at the VBM in the clean C-terminated surface are
contributed by the dangling bonds but also by back-bonds of the surface C-atoms 
(which, due to symmetry, also involve $p_{z}$ and $p_{y}$ combinations).
In the Si-terminated case, the larger adatom height from the surface favors an
overlap between Ni-$d_{3z^2-r^2}$ and Si-$p_{z}$ states, leaving the
$p_{y}$ component almost  not affected by the bond.

\subsection{Hollow sites}
Due to the symmetry of the four-fold adsorption site and the small vertical
distance from the surface atoms, the Ni states involved in the bonding have
mainly in-plane $d_{xy}$ and $d_{x^2-y^2}$ symmetry, as is evident from 
Fig.\ref{dos_hollow_si_c_term}.
In fact, we find a strong hybridization of $d_{xy}$-type states with the
Si- and C- $p$ states in the Si- and C-terminated 
surface, respectively; as in the bridge sites discussed previously,
these states form a bonding-antibonding configuration. In the C-terminated 
surface,  the energy splitting is more pronounced due to the larger orbital overlap: 
in fact, the anti-bonding $d_{xy}$-$p_{x}$ peak appears at about 1 eV above
$E_{F}$.
In both terminations, moving from the bulk to the surface layer, we find
the appearance of a  feature with $p_z$ character at $E_{F}$, located
in the middle of the SiC bulk band-gap. In order to better bring out the
nature of this feature, we plot in Fig.\ref{pz_hollow}(a) the projected DOS on C
atoms belonging to different layers. We observe that this same feature has 
a significant spectral weight also 
in the sub-surface layers (C2 and C3 in Fig.\ref{pz_hollow}(a)). 
These peculiar states originate from 
a vertical non-bonding combination of Ni $d_{3z^2-r^2}$ and $p_z$ orbitals from 
the atoms underneath the Ni adatom; their spatial extension and non-bonding 
 character are shown in Fig. \ref{pz_hollow}(b) where the charge density 
 corresponding 
  to states within 1 eV below the Fermi energy is plotted. 
While this peak falls right at $E_{F}$ in the Si-terminated
surface, it is completely filled in the C-terminated surface.

\section{surface-core-level shift}
\label{cori}
The surface-core-levels shifts are particularly relevant from an experimental 
point of view, since they are directly accessible by  photoemission
spectroscopy and can give insights on the adatom position and surface
reconstruction. We calculate the surface-core-level shift as the difference
of the core-level eigenvalue corresponding to the surface atom ($e.g.$ 
Si-$2p_{3/2}$ and C-$1s$) and that of the central (bulk) layer. 
It is clear that within 
this method we neglect so-called final-state effects, while taking into account only initial-state 
effects. 
Now, it is a common procedure to associate the observed core-level shifts with 
charge
transfer between the atoms, in our case Si (C) and the Ni adatom; however, the 
variation of the core-level eigenvalues is strictly related to changes in the 
total potential on the atom core, which is mainly due to two different 
 contributions\cite{ossicini}: an
intra-atomic effect, due to charge transfer, and an inter-atomic
contribution, namely the Madelung potential, that acts in the opposite
direction. A measure of the charge 
transfer can be achieved by comparing the charge in the muffin-tin (MT) region 
of the atom 
considered and the one in the inner layer taken as reference.

Figure \ref{cores}  shows the results obtained for the structures analyzed.
Even if the definition of charge transfer is not unique, the results show
 a well defined trend, namely, a  charge transfer from surface Si (C) to 
Ni, while all the other atoms remain bulk-like.  The charge transfer is rather small ($\sim$ 0.1 to $\sim$ 0.2
 electrons) in all the structures considered. This is consistent with  
 the trend observed in the core-level-shifts for all the structures except for 
 the
 Si-terminated with Ni in the bridge-site. In this last case, we observe a 
 positive shift towards lower binding
 energy, even if the surface atom loses charge. It is evident that a direct 
 relationship between charge transfer and core-level shift is misleading in this 
 case. 
 The correct behaviour is recovered if we  substract the Madelung potential 
 from the core-level binding energy \cite{madel}.
 The result is shown in Fig.\ref{cores} (a)(dashed line). We see that after a
 proper accounting of the Madelung term the core-level shift follows the charge
 transfer trend; we point out that subtraction of the Madelung term in the 
 other structures does not change the overall trend.
 The calculated core-level shift to lower binding energy of the Si 2$p_{3/2}$
 state
 agrees with the trends observed by photoemission spectroscopy\cite{hoc}, 
 but we showed that this does not directly imply a positive charge transfer
 towards the surface
 Si atom.   
 As a general remark, we point out that the order of magnitude of the charge
 transfer\cite{ossicini} and of the core-level shift are consistent with the 
 covalent character discussed previously  for the bond between Ni and SiC 
 where a strong charge transfer is not expected.

\section{Conclusions}
\label{concl}
The highly precise FLAPW method \cite{flapw} was used to understand the 
adsorption of a 
Ni monolayer on the $\beta$-SiC(001) surface. The adsorption energy on both the 
Si- and C-
terminated surfaces is found to be quite high and differs only slightly  
between the different adsorption sites (bridge and hollow). We predict, 
therefore, that the
SiC(001) surface represents a strong reactive surface for Ni and could
therefore be  a  suitable substrate for the growth of metallic Ni and the 
formation of
Ni-silicides. We explained the high adhesion energy in terms of a strong
covalent Si- (C-)Ni bond. In all the surface geometries  investigated, the 
Si- (C-)$p$
orbitals are strongly hybridized with the Ni-$d$ states resulting in 
bonding-antibonding features and a consequent partial lowering of the 
Ni-$d$ density of states at $E_{F}$. The surface states are well localized in the surface region
 (Ni overlayer and the topmost Si (C) layer) for Ni in the 
 bridge-site,
 while a deeper penetration of the antibonding state of $p_{z}$-$d_{3z^2-r^2}$
 character into the SiC bulk is observed for Ni in the hollow-site.
We calculated the  surface-core-level shifts and found them in agreement with 
available
experimental data;  we also showed that the Madelung contribution should be
correctly taken into account to
relate the observed core-level-shifts to the charge-transfer.
\section{Acknowledgments}
Work at Northwestern University was supported by the U.S. Department of Energy 
(under grant No. DE-F602-88ER45372).

\begin{table}
\centering
\small
\caption{Equilibrium bond-length ($d_{Ni-X}$) between Ni and (Si, C) surface 
atoms, vertical distances between the Ni adatom from the surface plane
($h$) (in \AA) and adsorption energies (in eV) }
\vspace{5mm}

\begin{tabular}{p{3.cm}p{2.cm}p{2.cm}p{2.cm}}
Si-terminated  & Antibridge&Bridge&Hollow \\ 
\hline 
$d_{Ni-Si}$    &  2.18     & 2.16 &2.38   \\
$h$            &  1.55     & 1.52 &1.00      \\
$E_{ads}$      &  6.61     & 7.42  & 7.40   \\
\hline\hline
C-terminated   &           &      &       \\
$d_{Ni-C}$    &  1.82      & 1.78 & 2.22  \\
$h$            & 0.97     & 0.92  & 0.50   \\
$E_{ads}$      & 6.90     & 7.35  & 7.30   \\
\end{tabular}
\label{param}
\end{table}

\begin{figure}
\caption{Side and top view of SiC(001) ($1\times1$) showing the different
 Ni adsorption sites.} 
\label{struct}
\end{figure}

\begin{figure}
\caption{PDOS for the bridge adsorption site in the (a) Si and (b) C 
terminated surface. Upper panel: Ni $d$-projected
density of states; central panel: topmost surface Si (a) and C (b) layer, 
$s$ (dotted line), $p_{x}$ (bold line), $p_{y}$ 
(dashed line), $p_{z}$ (thin line) contributions; lower panel: bulk Si (a) and 
C (b) layer s (dashed line) and p (thin solid line) contributions.}
\label{dos_bridge_si_c_term}
\end{figure}

\begin{figure}
\caption{Charge density plot for the (a) bonding  and (b) antibonding  peak for 
the Si-terminated surface with Ni in the bridge adsorption site. The charge
densities are constructed taking into account only states belonging to the 
peak at -3 eV (for the bonding) and at 1eV (for the antibonding state).}
\label{bond_ant}
\end{figure}

\begin{figure}
\caption{PDOS for the hollow adsorption site. Symbols as in Fig. 
\ref{dos_bridge_si_c_term}}
\label{dos_hollow_si_c_term}
\end{figure}

\begin{figure}
\caption{Panel (a): $p_{z}$-like contribution to the density of states 
from carbon atoms at different slab layers; panel (b): charge density plot 
due to 
states up to 1 eV below $E_{F}$. Both panels refer to the hollow-site geometry,
labels C1 up to C4 refer to sites from surface down to bulk layers.}
\label{pz_hollow}
\end{figure}

\begin{figure}

\caption{Surface-core level shifts for Si 2$p_{3/2}$ 
for (a) Ni in the bridge-site  and (c) Ni in the hollow-site  
and total charge inside the Si-MT (insets therein).
Panel (b) and (d): C-1$s$ core level shift and total MT charge (insets) for
the bridge and hollow site, respectively.}
\label{cores}
\end{figure}

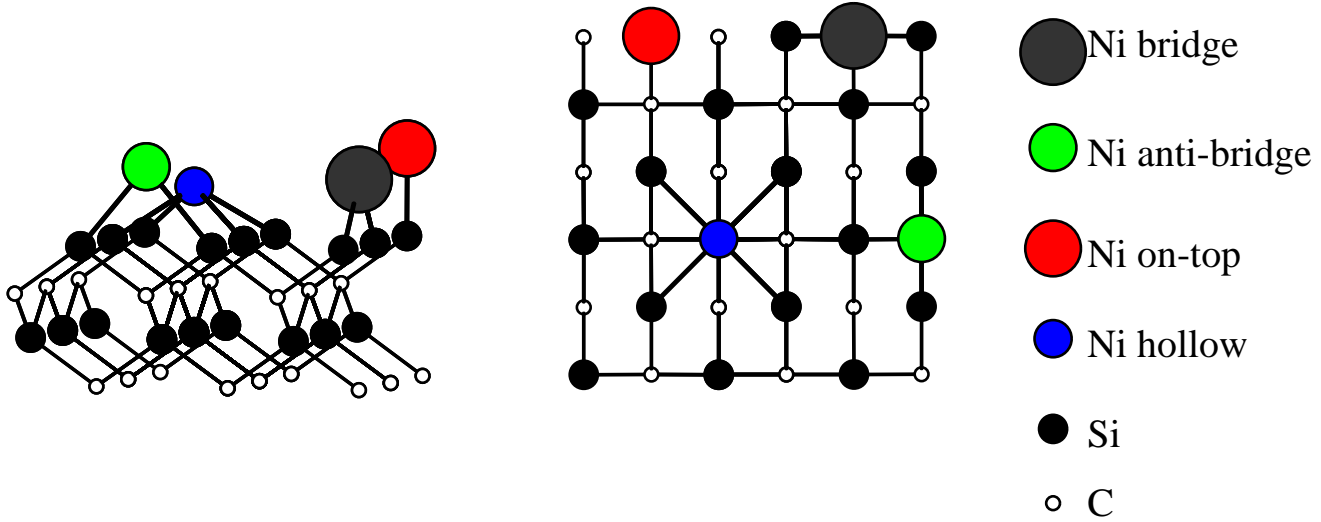
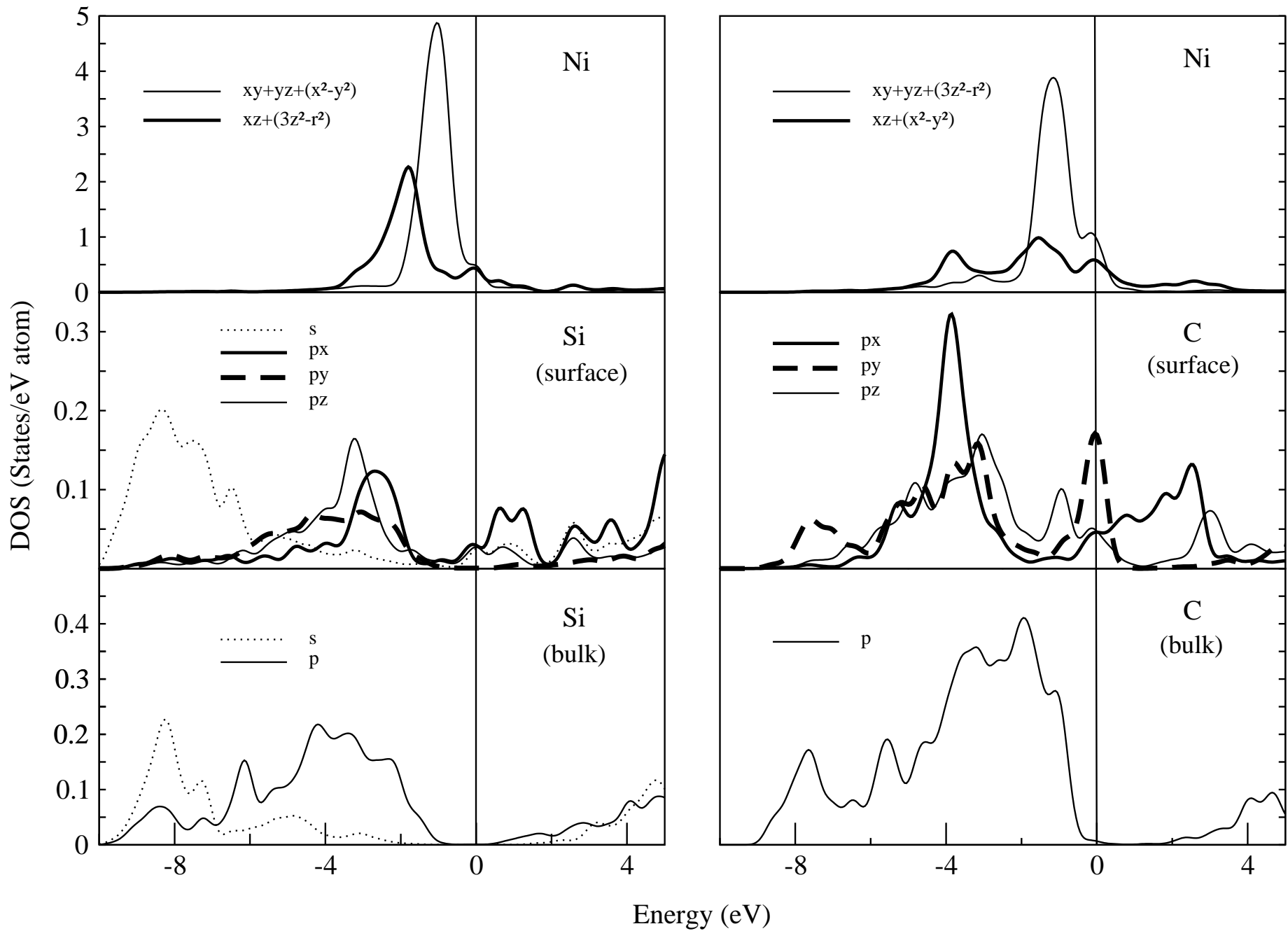
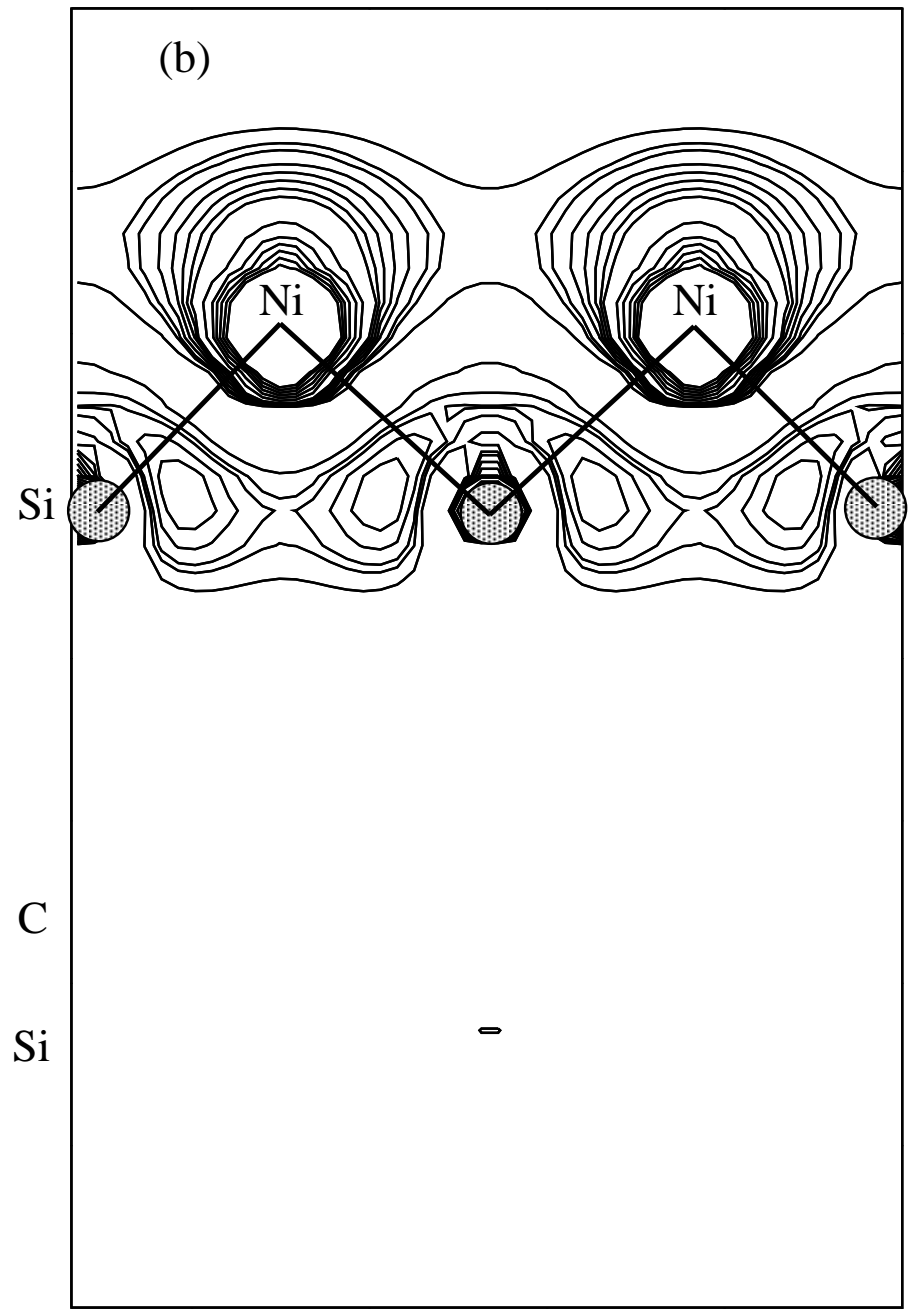
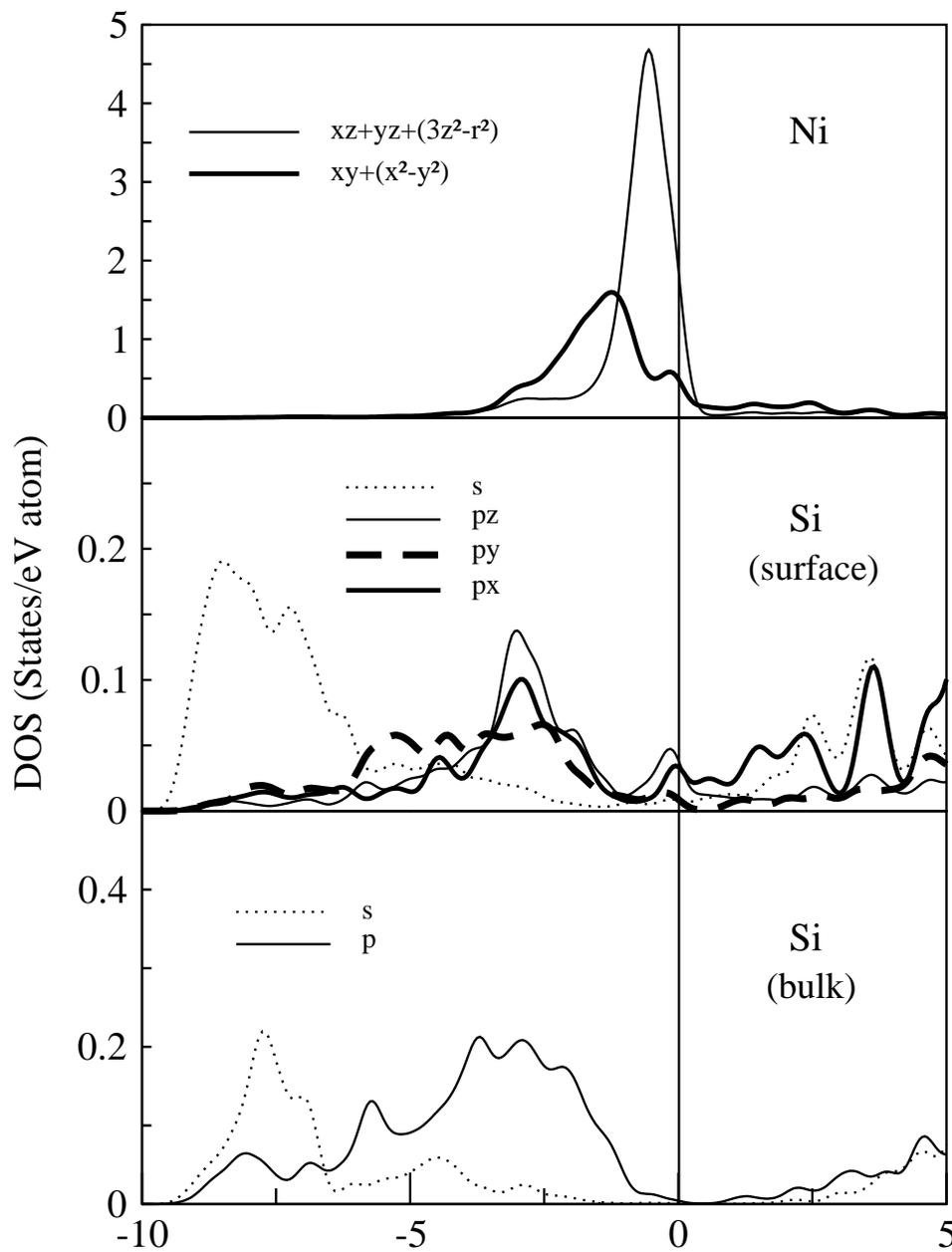
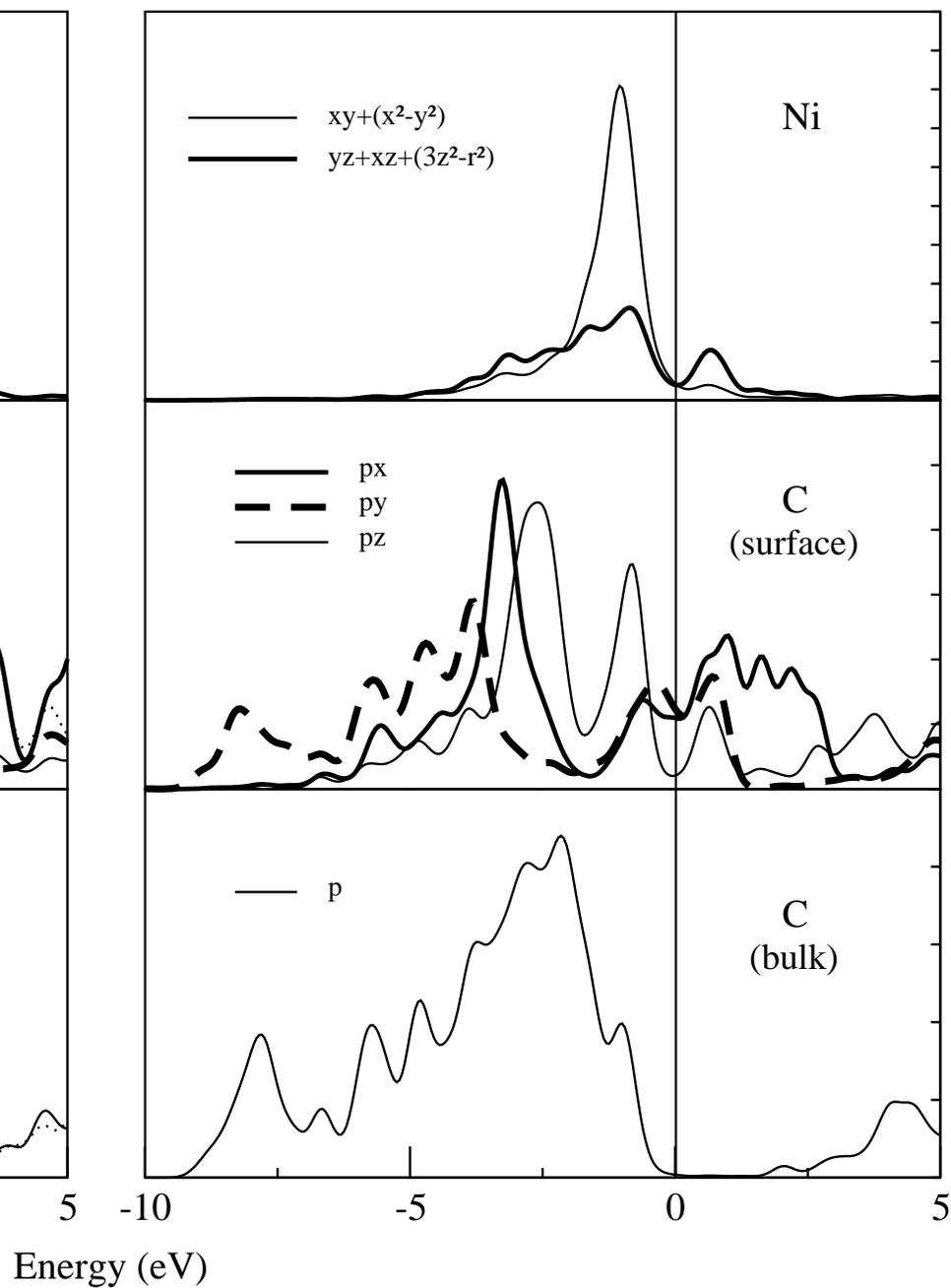
\end{document}